# Beyond Publication-Bias Detection:

# Estimating Bias under Uncertainty in Heterogeneous Literatures

Ulrich Schimmack

University of Toronto, Mississauga

**Authors' Note**

Ulrich Schimmack https://orcid.org/0000-0001-9456-5536. Correspondence concerning this article should be addressed to Ulrich Schimmack, Department of Psychology, University of Toronto, 3359 Mississauga Road, Mississauga, ON L5L 1C6, Canada. Email: ulrich.schimmack@utoronto.ca.

**Abstract**

Meta-analysts routinely test for publication bias, but a nonsignificant test is inconclusive because it may be a Type II error. In a large factorial simulation, I show that publication-bias tests have low power even with 1,000 studies once effect sizes are heterogeneous and the data do not meet the model's assumptions. I argue that the goal should shift from detecting bias to estimating it with confidence intervals. A confidence interval does not only bound the hypothesis that bias is absent; its upper bound also indicates how much bias remains compatible with the data. When the interval is wide, the upper bound does not exclude a large amount of bias, even if the formal test is nonsignificant. I then show that confidence intervals from step-function selection models are too narrow, covering the true amount of bias only about half the time. In contrast, z-curve's confidence interval — obtained by transforming the expected discovery rate into a selection-weight parameter — has nominal coverage. Z-curve is therefore a valuable tool for examining publication bias in meta-analyses, especially when heterogeneity is high. I illustrate the implications of bias estimation with a meta-analysis of social priming and a set of applied-psychology meta-analyses.

**Beyond Publication-Bias Detection:**

**Estimating Bias under Uncertainty in Heterogeneous Literatures**

Publication bias distorts the scientific record. If published results are selected to favor statistical significance, significance loses its significance (Sterling, 1959; Sterling et al., 1995). In the worst-case scenario, a literature can consist largely of false-positive findings when nonsignificant results remain unpublished (Rosenthal, 1979). The prevalence of significant results also differs across fields. Fanelli (2010, 2012) showed that positive results are more common in some disciplines than others and that negative results have become less common over time. The failure to replicate significant results has raised awareness that selective publishing of significant results undermines the credibility of original research articles (Open Science Collaboration, 2015).

Publication bias also undermines the credibility of meta-analyses. For example, meta-analyses of extrasensory perception studies have sometimes produced apparently positive evidence for ESP, but a more plausible explanation is selective reporting or publication bias (Galak et al., 2012; Schimmack, 2012). To address this problem, several statistical methods have been developed and evaluated (Carter et al., 2019; Hong & Reed, 2021; Renkewitz & Keiner, 2019).

A major limitation of this literature is that evaluations have focused primarily on the recovery of the mean population effect size. This focus is defensible for meta-analyses of homogeneous literatures, such as direct-replication studies or clinical trials that estimate the effect of a specific treatment on a specific outcome. However, the same methods are often applied to conceptual-replication literatures in which studies differ in samples, procedures, measures, outcomes, and theoretical instantiations. In such literatures, the mean effect size can be a poor summary of the distribution of effects, and heterogeneity is not a nuisance parameter but a central feature of the evidence.

Moreover, correction for publication bias often requires strong assumptions about unknown quantities. This explains why no method performs uniformly well across different conditions in simulation studies (Carter et al., 2019). Evaluating bias correction while neglecting bias identification puts the cart before the horse. Evidence of bias is valuable information, even if it is not possible to obtain corrected estimates from a biased set of studies.

Relatively few studies have examined the power of publication-bias tests to detect the presence of bias, and the results are not encouraging (Hedges & Vevea, 1996; Renkewitz & Keiner, 2019). Hedges and Vevea (1996) evaluated the performance of their step-function selection model with different amounts of selection bias, distributions of population effect sizes, and study-set sizes. The largest set size was 100 original studies, and power to detect extreme bias was below 25%. Renkewitz and Keiner (2019) compared several bias-detection methods and found that some methods could detect bias when all nonsignificant results were missing. However, when as little as 10% of nonsignificant results were included in the meta-analysis, power declined sharply, especially in the presence of heterogeneity. These simulation results suggest that bias detection in meta-analysis with fewer than 100 studies is practically impossible.

A key limitation of previous simulation studies is their focus on relatively small sets of studies. This focus is appropriate for meta-analyses of direct replications or narrowly defined clinical literatures, but meta-analyses of conceptual-replication literatures can include far more than 100 studies because they address broad research questions rather than one specific paradigm. The first contribution of this article is therefore to examine the performance of bias-detection methods when larger sets of studies are available.

The second contribution is to examine the robustness of bias-detection methods across different distributions of population effect sizes and different selection mechanisms. Some methods can perform well when the simulation matches their assumptions but perform poorly when those assumptions are violated (Brunner & Schimmack, 2020; Hedges & Vevea, 1996). This issue is often obscured in simulation designs that use a single population distribution or a single selection function. For example, normally distributed population effects favor methods that assume normality. When the true distribution of effects is skewed, bimodal, or contains a mixture of null and non-null effects, performance declines (Hedges & Vevea, 1996).

The third contribution is to shift the focus from binary tests of publication bias to estimation of the possible amount of bias. When power is low, nonsignificant bias tests cannot distinguish between the absence of bias and the failure to detect it (false negative results). Estimation with confidence intervals overcomes this limitation of significance tests. The critical information is the upper bound of the confidence interval. A low value implies that publication bias is small and practically negligible, whereas a high value suggests that publication bias could severely bias effect size estimates. To serve this purpose, it is important to validate the coverage of confidence intervals. That is, confidence intervals with upper bounds that are too low do not properly inform researchers about the risk of publication bias.

There are various measures of publication bias. In regression models, the slope of the regression of effect size estimates on sampling error quantifies bias. However, the problem is that other factors can influence this correlation and that the slope is not a direct measure of the amount of missing studies. Weight-function selection models quantify bias with weights for steps of p-values. A weight for nonsignificant results of .5 implies that 50% of nonsignificant results are missing. This information is more direct than regression slopes, but the focus on nonsignificant results means that the amount of bias in the full dataset can vary. If most studies are significant, even 50% missing nonsignificant results have no notable influence on effect size estimates. However, if most results are nonsignificant, removing 50% of them, can dramatically inflate effect size estimates. A better way to quantify the amount of bias is the amount of missing studies from the entire dataset. For example, if 100 studies have an average power of 50%, and 50% of nonsignificant results are missing, we are left with 50 significant results and 25 nonsignificant results. The observed discovery rate (i.e., percentage of significant reported results) is now 50/75 = 66%, an inflation of 16 percentage points over the true discovery rate of 50%. There is no need to choose between the two estimands because they are merely different expressions of the same information relative to different denominators. This also makes it possible to compare models that use different estimands directly to each other.

In short, this article examines a fundamental problem in psychological science. Researchers rarely conduct direct replications, and single studies often have low power. Meta-analysis is often proposed as a solution, but meta-analyses of conceptual-replication literatures frequently have substantial heterogeneity. Publication bias can inflate effect-size estimates in such literatures, yet bias may be difficult to detect. As a result, many meta-analyses may report inflated effect-size estimates and provide misleading evidence for psychological phenomena. Although these concerns have gained visibility during the replication crisis, the specific problem of low power for detecting publication bias in heterogeneous literatures has received insufficient attention (Renkewitz & Keiner, 2019).

## Selection of Methods

Broadly speaking, existing methods detect publication bias from three types of diagnostic information: (a) small-study effects, (b) excess significance or power discrepancies, and (c) discontinuities in the distribution of test results around critical values, such as the conventional significance threshold. Tests within each group rely on similar assumptions and can therefore be discussed as families of methods rather than one by one. These assumptions also make it possible to exclude some methods from primary consideration because they are unlikely to be competitive as bias-detection methods under realistic conditions.

**Small Study Effects**

The earliest and most widely used tests of publication bias are small-study-effect tests, including Egger's regression test (Egger et al., 1997). These tests examine whether effect-size estimates are larger in studies with larger standard errors than in studies with smaller standard errors. One possible explanation for such a relation is publication bias. When smaller studies have larger standard errors, they require larger observed effects to reach statistical significance. If significant results are preferentially published, small studies will therefore tend to show larger published effects than large studies.

However, this logic has several limitations (Sterne et al., 2011). First, small-study-effect tests do not identify publication bias directly. They detect a pattern that is merely consistent with publication bias. The same pattern can arise for other reasons, including genuine differences between small and large studies. This limitation is especially problematic in heterogeneous literatures, where sample size, design, population, manipulation strength, measurement reliability, or outcome type may be correlated with the population effect size.

Second, previous simulation studies have shown that regression-based methods can have low power and inflated Type I error rates in heterogeneous meta-analyses (Sterne et al., 2011; Lin & Chu, 2018). Moreover, heterogeneity increases the possibility that factors other than publication bias produce correlations between effect sizes and sampling error. Even proponents of regression-based methods have expressed concerns about their performance when heterogeneity is high (Stanley, Carter, & Doucouliagos, 2018). Stanley and colleagues later developed an excess-significance test designed to address some of these limitations, and this test was included in the present study (Stanley, Doucouliagos, Ioannidis, & Carter, 2021).

Another reason not to include regression-based methods as primary competitors is that they require notable variation in sampling error with at least some studies with small sampling error that have relatively little bias in effect size estimates. This requirement makes these models less suitable for meta-analyses of research areas in which large studies with small standard errors are expensive, rare, or unavailable.

Last, it is not sufficient to detect bias; it is also important to quantify the amount of bias (Lin & Chu, 2018). Regression methods can estimate small-study effects or bias-corrected mean effects, but their coefficients are not directly comparable to more intuitive measures of selection bias, such as the proportion of nonsignificant results that are missing or the discrepancy between the observed and expected frequency of significant results.

**Excessive Significance and Power Discrepancies**

A second class of bias tests uses statistical power to detect publication bias. In this literature, however, the term power is often used more broadly than in conventional a priori power analysis for sample-size planning (Bartoš & Schimmack, 2022; Brunner & Schimmack, 2020; Soto & Schimmack, 2026). Classical power is conditional on the null hypothesis being false and on a specified population effect size being the true effect size. This conditional definition is useful for planning individual studies, but it is less suitable for heterogeneous meta-analytic literatures in which studies may differ in their population effects and some tests may concern true null hypotheses.

For this reason, Bartoš and Schimmack (2022) introduced the term expected discovery rate (EDR) for the average unconditional probability that a set of studies will produce statistically significant results. Unlike classical power, the EDR does not condition on all null hypotheses being false. In this article, I therefore use the more precise term EDR, although the broader literature often uses the term power more loosely. EDR-based bias tests compare the observed discovery rate (ODR), defined as the proportion of reported tests that are statistically significant, with the EDR. If the ODR is larger than the EDR, the published record contains more significant results than expected under the model. This discrepancy provides evidence that statistically significant results may have been selectively reported or published.

EDR-based bias tests have several advantages over small-study-effect tests. First, they do not rely on correlations between effect-size estimates and sampling error. As a result, they are less vulnerable to confounding between study precision and true-effect heterogeneity, which is the central problem for regression-based funnel-asymmetry tests. Second, EDR-based tests do not require meaningful variation in sampling error. Small-study-effect tests require variation in standard errors because they detect bias from the association between effect size and precision. EDR-based tests can be applied even when all studies have the same sample size and sampling error.

A third advantage is that some EDR-based methods make weaker assumptions about the distribution of nonsignificant results. This is important because the nonsignificant part of the evidence is often precisely the part that is missing or selectively reported. Step-function selection models, for example, model the relative publication probability of significant and nonsignificant results and typically assume a particular distribution of population effects. These assumptions can be useful when approximately correct, but they can also make bias tests sensitive to model misspecification. By contrast, methods that estimate the EDR from the distribution of significant results are less dependent on the distribution of nonsignificant results.

A final advantage of EDR-based tests is that they quantify the amount of excess significance. In a large dataset, a local discontinuity test may become statistically significant even when the discontinuity around p = .05 is small and has limited influence on effect-size estimates. By contrast, the comparison between ODR and EDR estimates the extent to which the observed proportion of significant results exceeds the expected proportion. This makes it possible to evaluate not only whether the evidence is suspicious, but also how much selection may have distorted the published record.

The main challenge for EDR-based tests is to estimate the true EDR of a set of studies. The Test of Excess Significance (TES) was the first EDR-based test (Ioannidis & Trikalinos, 2007). TES uses the average effect size of a meta-analysis to estimate the power of individual studies. This approach has two important limitations. First, the EDR estimate is inflated if the meta-analytic mean itself is inflated by publication bias. Second, TES treats all studies as if they share the same true effect size. In this sense, TES does not model a distribution of population effects; it assumes a degenerate distribution with no heterogeneity. This assumption is implausible in meta-analyses of conceptual replications, where studies often differ in samples, procedures, measures, outcomes, and theoretical instantiations (Schimmack, 2012).

Stanley et al.'s (2021) Publication Selection Significance Test (PSST) relaxed the equal-effect assumption by allowing for heterogeneity in effect sizes and proposed a correction for inflated Type I error rates when studies are heterogeneous. However, PSST still relies on a meta-analytic estimate of the average effect to compute the expected number of significant results. This makes the test sensitive to the same general problem as TES: when the meta-analytic estimate is inflated by selective reporting, the estimated EDR may also be inflated, reducing power to detect bias.

The Incredibility Test estimates the EDR from observed power estimates for individual studies (Schimmack, 2012). However, these observed power estimates are themselves functions of the observed effect sizes. If observed effects are inflated by publication bias, the estimated EDR is also inflated, which lowers the power of the test to detect bias. One response to this problem is to adjust for inflation by using the discrepancy between ODR and estimated EDR (Schimmack, 2016). The R-Index is defined as EDR - (ODR - EDR), or 2 × EDR - ODR. It is more powerful than the Incredibility Test, but it may also have higher false-positive rates. A key limitation of observed-power approaches is that observed power distributions are not normal and that average observed power can be a poor measure of average true power (Yuan & Maxwell, 2005).

**Discontinuity Tests**

The caliper test was introduced by Gerber and Malhotra (2008a, 2008b) as a local test of discontinuity around the significance threshold. As Rosenthal (1979) famously noted, “God loves the .06 nearly as much as the .05.” Thus, in the absence of selection, one would not expect a sharp discontinuity in the distribution of p-values immediately around the conventional significance criterion. The caliper test turns this expectation into a bias test by comparing the frequency of p-values just below and just above a criterion value, typically $p = .05$. Publication bias or selective reporting is inferred when there are significantly more p-values just below .05 than just above .05.

The main advantage of the caliper test is that it is local and relatively assumption-light. It does not require estimation of a population effect-size distribution, a mean effect, a heterogeneity parameter, or an expected discovery rate. Instead, it relies on the weaker assumption that, in the absence of selection, the distribution of p-values should be locally smooth around the significance threshold. A sharp excess of p-values just below .05, relative to just above .05, is therefore evidence that the reporting process changes at the significance criterion. If nonsignificant results immediately above .05 are underrepresented relative to significant results immediately below .05, the pattern is consistent with publication bias, selective reporting, or p-hacking around the threshold.

The limitation is that this strength is also a weakness. The caliper test detects only local discontinuities. It can reveal selection around a known threshold, but it does not estimate the overall magnitude of publication bias in the literature. A literature may show strong selection for significance without producing a large local pile-up around .05, especially if many selected results are well below the threshold or if p-values are reported only coarsely. Conversely, in large datasets, very small local discontinuities can become statistically significant even if they have limited implications for the credibility of the meta-analytic results.

A second limitation concerns the width of the caliper. The assumption of local smoothness is most plausible in very narrow intervals around the significance threshold, but narrow intervals often contain few observations and therefore provide low power. Wider intervals increase the number of observations, but they also weaken the local-comparison logic. When the underlying p-value distribution is not approximately smooth over the chosen interval, differences between the two sides of the threshold may reflect the shape of the evidential distribution rather than selection. This can reduce power when most studies have low power and may increase Type I error when most studies have high power. Thus, the caliper test may perform best when the underlying distribution around the significance threshold is approximately smooth, but its performance can deteriorate when this condition is violated (Schneck, 2017).

Other methods model discontinuities in the selection process rather than testing only a local discontinuity in the observed p-value distribution (Andrews & Kasy, 2019; Hedges, 1992; Vevea & Hedges, 1995; Vevea & Woods, 2005). Although the basic idea is old, applications of these models in psychology became more common after the replication crisis (Carter et al., 2019). Step-function selection models can be understood as extensions of random-effects meta-analysis that add an explicit model of selective reporting. A random-effects meta-analysis estimates the mean and variability of population effect sizes under the assumption that all relevant results are observed. A step-function selection model allows for the possibility that results in some p-value regions are less likely to be observed than results in other regions.

For example, the simplest version is the three-parameter selection model (3PSM). It estimates the mean effect, heterogeneity, and one selection parameter that captures the relative probability that nonsignificant results are observed compared with significant results. The step is typically placed at the conventional significance threshold, corresponding to $z = 1.96$ for sign-normalized two-sided tests. More flexible models add additional steps, allowing selection probabilities to differ across multiple p-value regions.

Step-function selection models are relatively assumption-heavy. To estimate publication bias, they must specify both a latent distribution of population effect sizes and a selection process. Because these models were developed largely for applications to relatively homogeneous literatures, the latent distribution is usually modeled as normal. This assumption may be reasonable and robust when heterogeneity is small, but it can become problematic in heterogeneous literatures (Hedges & Vevea, 1996).

A second strong assumption concerns the selection process. Step-function models assume that publication bias acts uniformly on all results within a given p-value interval. For example, in a one-step model, all nonsignificant results receive the same selection weight. This assumption implies that selection reduces the number of nonsignificant results but preserves their shape within the nonsignificant region. If selection operates differently across the nonsignificant range, for example if p = .06 results are more likely to be reported than p = .60 results, or if large-sample null results are reported for different reasons than marginally nonsignificant results, the estimated selection weights and latent effect distribution may be biased.

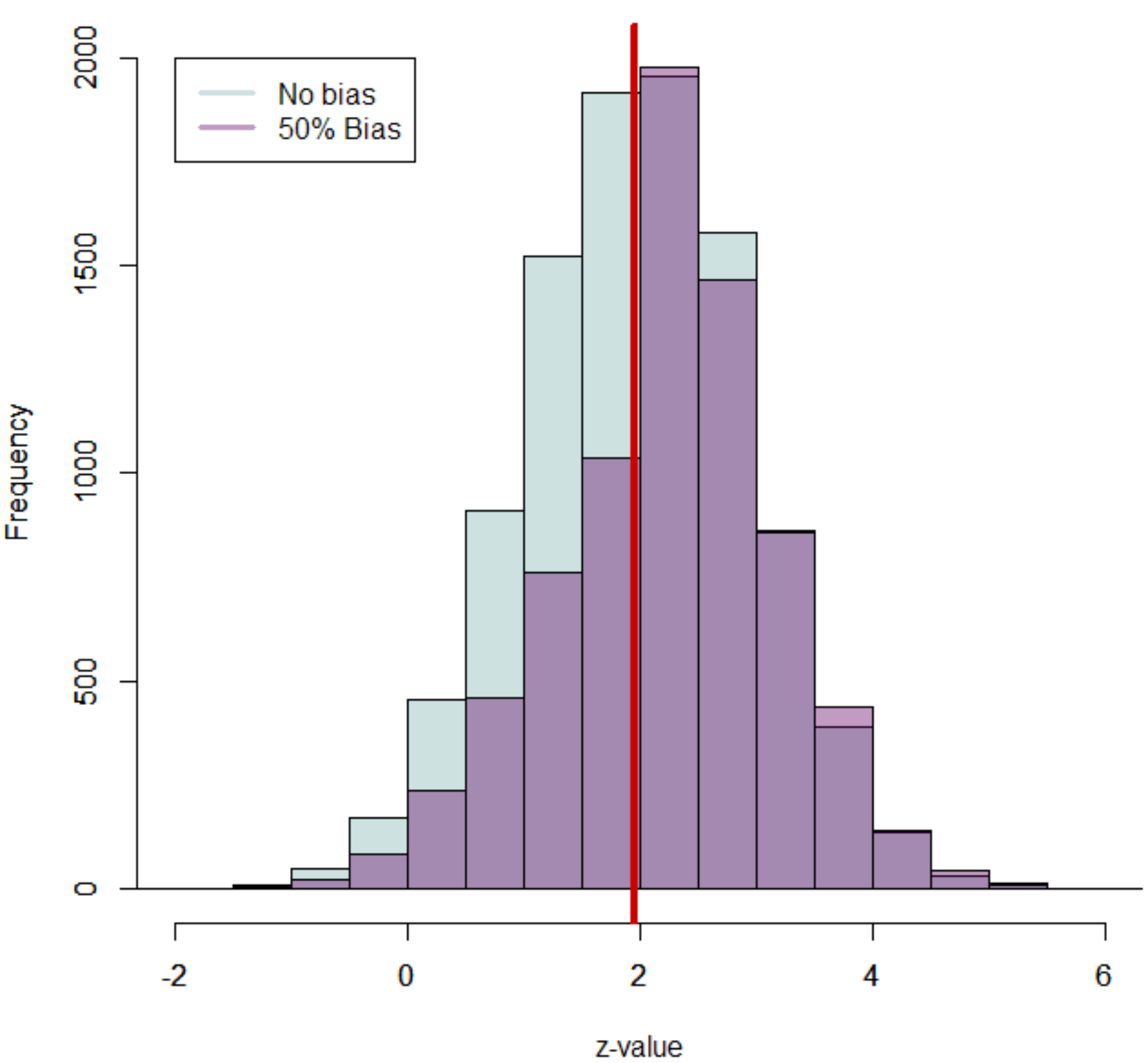

Figure 1

Histogram of simulated z-values with and without publication bias (normal distribution)

Figure 1 illustrates this implication with a simple simulation. The simulation assumes homogeneous effects, equal sample sizes, and a noncentrality parameter of 2, corresponding to approximately 50% power for a two-sided test at $\alpha = .05$. Publication bias was simulated by uniformly dropping 50% of the nonsignificant results while retaining all significant results. As expected, the selected distribution shows a discontinuity at the significance threshold. However, the nonsignificant results are not flattened or reshaped. They remain a downweighted continuation of the no-bias distribution. This is a direct consequence of the step-function assumption: uniform selection within the nonsignificant region changes the frequency of nonsignificant results, but not their distributional shape.

Applying the selection model of Vevea and Woods (2005) to these simulated data shows that the model performs well when its assumptions are approximately satisfied. It recovers the true mean effect and heterogeneity accurately, and it detects publication bias: the 95% confidence interval for the weight assigned to nonsignificant results is well below 1, estimated weight = .54, 95% CI [.51, .59]. The model also provides accurate estimates of the simulated mean, d = .41, 95% CI [.40, .42], and heterogeneity, $\tau = .04$.

A second scenario examined the robustness of the model against violation of its assumptions. In the simulation, the population mean was d = 0 with moderate heterogeneity, tau = .4. The simulated distribution was a beta distribution with shape parameters alpha = 1 and beta = 10 with mean 0 and standard deviation $\tau = .4$. Selection was simulated as a graded function of power rather than as a step function at statistical significance. The overall amount of selection was again calibrated to remove 50% of the nonsignificant results.

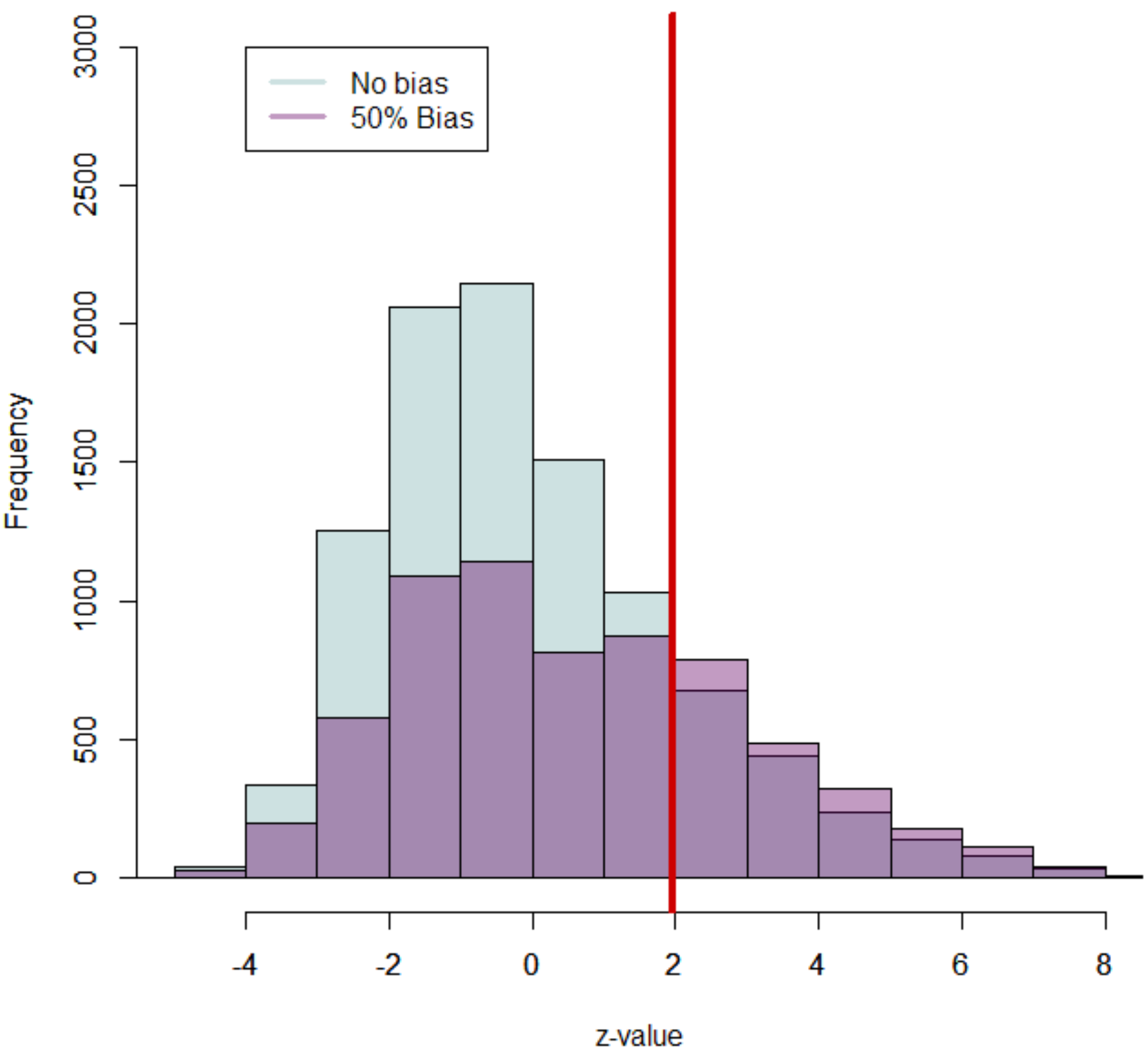

Figure 2

Histogram of simulated z-values with and without publication bias (non-normal distribution)

Figure 2 shows the distribution of z-values. In this scenario, the step-function model fails to detect publication bias in the correct direction. In fact, the estimated weight points in the opposite direction: nonsignificant results are estimated to be more likely to be observed than significant results, weight = 1.20, 95% CI [1.11, 1.30]. In this simulation, publication bias inflates the unadjusted estimate of the population effect size to d = .13, 95% CI [.12, .14]. Because the step-function model estimates reverse selection, it does not correct this bias and instead produces a slightly larger adjusted estimate, d = .17, 95% CI [.14, .18]. The model also overestimates heterogeneity, $\tau = .48$. This example illustrates the risk of assumption-heavy selection models. When the latent effect distribution and selection process are misspecified, the model can misinterpret distributional shape as selection, fail to detect actual publication bias, and produce a correction that moves the estimate farther from the truth. Two wrong assumptions do not make a right correction.

**Z-Curve**

A more recent power-based method is z-curve 2.0 (Bartoš & Schimmack, 2022). Z-curve was explicitly designed for heterogeneous sets of studies, including literatures in which studies address different research questions, use different designs, and vary widely in sample size and effect size. This makes z-curve suitable for evaluating the statistical credibility of results published in journals, research areas, or broad theoretical literatures, rather than only narrowly defined meta-analyses of a single effect (Schimmack, 2020; Schimmack & Bartoš, 2023; Soto & Schimmack, 2025). A key advantage of z-curve is that it does not require a model for the distribution of population effect sizes. It can also accommodate mixtures of true and false null hypotheses (Brunner & Schimmack, 2020).

Z-curve 1.0 used a finite mixture model to correct for the inflation of observed power when studies are selected for significance. Z-curve 2.0 extended this framework to estimate the expected discovery rate (EDR) using only statistically significant results. The model assumes a simple selection process: statistically significant results are likely to be observed, whereas nonsignificant results may be missing. To infer the expected number of nonsignificant results, z-curve uses the estimated mixture weights of components with different levels of power. A high proportion of just-significant results implies a large contribution from low-powered studies, because high-powered studies tend to produce both moderate and large significant z-values rather than clustering near the significance threshold.

Simulation studies showed that z-curve 2.0 can estimate the EDR with systematic bias in some specific situations. To address this problem the confidence intervals were adjusted by 5 percentage points to ensure nominal coverage in all scenarios (Bartoš & Schimmack, 2022). The main limitation is that confidence intervals can be wide in some settings, especially with small sets of studies or weak information in the significant z-value distribution (Soto & Schimmack, 2026). However, z-curve's performance as a bias-detection tool has not been systematically compared with other methods. As the power of other methods is also modest, z-curve is a viable contender.

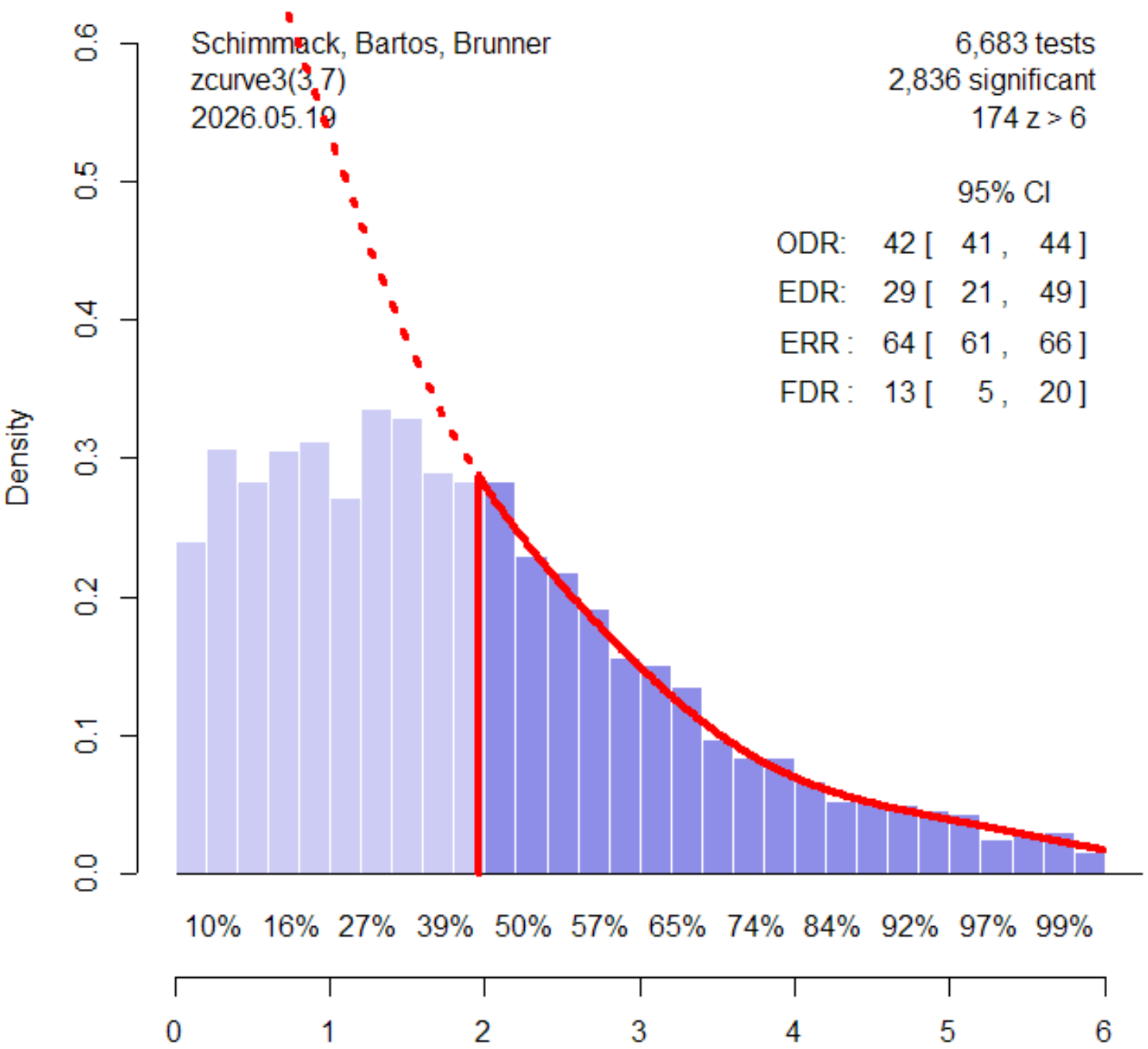

Figure 3

Z-curve plot of simulated data with beta(1, 10) effect distribution and graded selection bias

Figure 3 shows a z-curve plot based on the same data used in Figure 2. The main difference is that z-curve ignores the sign of z-values and models the distribution of sign-normalized significant z-values. Z-curve does not model the full distribution of observed effect sizes. Instead, it uses the significant z-values to estimate the mixture weights of components with different levels of power. These weights are then used to predict the expected distribution of nonsignificant results. In Figure 3, the observed significant z-values are shown in dark purple, and the fitted z-curve is shown by the solid red line. The observed nonsignificant z-values are shown in light purple, and the predicted nonsignificant distribution is shown by the dashed red line.

In this example, z-curve indicates some publication bias because the fitted model predicts an increasing density of nonsignificant results below the significance threshold, whereas the observed distribution drops sharply. The estimated amount of bias is the difference between the observed discovery rate and the expected discovery rate. In this simulation, ODR − EDR = 13 percentage points. However, even with more than 2,000 significant z-values, the confidence interval for the EDR estimate is wide and includes the ODR. To formally test publication bias and to compare z-curve with other methods, z-curve 3.0 uses bootstrapped difference scores of ODR and EDR to test publication bias. The 90% confidence interval ranges from -.05 to +.20. Thus, the data are compatible with publication bias, but the power to reject the null hypothesis of no bias is limited.

This example illustrates an important distinction between bias detection and bias estimation. Z-curve may fail to produce a statistically significant bias test when the significant z-value distribution is weakly informative, especially under high heterogeneity. However, the resulting confidence interval still provides useful information about the range of bias that remains compatible with the data.

**Other Methods**

P-curve was not included because it is primarily an evidential-value and bias-correction method, not a direct publication-bias detection test (Simonsohn, Nelson, & Simmons, 2014). P-curve analyzes only statistically significant results; nonsignificant results are ignored even when they are available. As a result, p-curve does not directly test whether nonsignificant results are underrepresented in a dataset. This makes it less suitable for the present study, which focuses on detecting publication bias. In addition, p-curve is most appropriate for relatively homogeneous sets of studies that test the same underlying effect. Its estimates can be severely biased when heterogeneity is substantial (Schimmack & Soto, 2026).

P-uniform was not included for similar reasons. The original p-uniform method uses only statistically significant results and assumes effect-size homogeneity, although it does include a publication-bias test (van Assen, van Aert, & Wicherts, 2015). The newer p-uniform* method was developed to address heterogeneity by using both significant and nonsignificant effect sizes and by estimating between-study variance (van Aert & van Assen, 2025). However, p-uniform* is primarily a correction method for estimating the mean effect size and heterogeneity under publication bias, not a direct test of whether nonsignificant results are underrepresented. The package documentation describes p-uniform as including a publication-bias test, whereas p-uniform* is described as an extension for estimating the average effect size and between-study variance. Thus, including original p-uniform would test a method known to be misspecified for the present heterogeneous design, whereas including p-uniform* would shift the comparison from bias detection to correction.

Another recent method is Robust Bayesian Meta-Analysis (RoBMA; Bartoš et al., 2023; Maier et al., 2023). RoBMA was not included in the main comparison because it is a model-averaging framework rather than a distinct bias-detection principle. Its evidence for publication bias is based on averaging across familiar components, including regression-based small-study-effect models and step-function selection models. In the present simulation, regression-based components are expected to add little information because sample sizes were fixed within conditions and therefore provided little or no variation in sampling error. Preliminary simulations confirmed this expectation: the step-function selection component dominated the Bayesian model average, whereas regression components received little weight.

RoBMA also averages across different step-function specifications. This is useful when the selection process is unknown, but it is not necessarily advantageous in the present simulation because the selection process was known and matched by the Vevea–Hedges specification used in weightr. Thus, the weightr model was given the optimal step specification for the simulated selection mechanism. Under these conditions, RoBMA would be expected to perform similarly to, or worse than, the correctly specified step-function model while requiring substantially more computation. For this reason, RoBMA was not included in the full simulation design.

**Simulation Design**

The results of simulation studies depend strongly on the selected scenarios. One problem is that simulations can be designed in ways that closely match a method's assumptions. Under these favorable conditions, a method may perform well even if it is fragile under more realistic conditions. For example, step-function selection models can outperform z-curve when true effect sizes are normally distributed, but their performance deteriorates when this assumption is violated (Brunner & Schimmack, 2020). A second problem is that all methods may fail under unusual or easily diagnosed conditions. Focusing on such edge cases does not provide a useful assessment of a method, especially if competing methods also fail under the same conditions.

To cover a broad range of plausible scenarios, the present study used a simulation design previously used to validate confidence-interval coverage for z-curve's EDR estimates (Bartoš & Schimmack, 2022). The design varied seven factors: (a) the number of significant studies (k = 100, 300, 1000), (b) the mean effect size (d = 0, .2, .4, .6), (c) heterogeneity in effect sizes (tau = 0, .2, .4, .6), (d) sample size (N = 50, 100, 200), and (e) the proportion of significant results that were false positives, that is, the false discovery rate (FDR, 0, .2, .4, .6). In addition, the study simulated three types of publication bias. The no bias condition was used to estimate type I error rate. A simple step condition fit the assumptions of the step-function model and serves as a sanity check. The step-function model should do well in this condition because the data match the assumption of the model. The third form of selection bias used graded selection. This condition tests the robustness of step-function selection models. The last factor varied the distribution of studies that tested a true hypothesis. A normal distribution is often used and fits the assumptions of step-function models. A truncated normal was used because it seems plausible that some literatures mostly test positive effects. A beta distribution was used to model literatures where most effects are small, but some studies have large effects. This design has 5,184 conditions.

**Publication Bias Tests**

1. The Incredibility Test (ICT; Schimmack, 2012). This test uses mean observed power as an estimate of the expected discovery rate. Publication bias is inferred when the observed discovery rate is significantly higher than the expected discovery rate. The advantage of this approach is that it does not assume a common population effect size. The limitation is that observed power is inflated when publication bias is present, which can make the test conservative.

2. The Replication Index (RI; Schimmack, 2014). The Replication Index corrects mean observed power for inflation when publication bias is present. The formula is EDR − (ODR − EDR), or equivalently 2 × EDR − ODR. For example, if mean observed power is .75 and only significant results are observed, the Replication Index estimate is 2 × .75 − 1 = .50. This correction can make the test more powerful than the Incredibility Test, but it can also increase Type I error rates if the correction overadjusts and the EDR is underestimated.
3. The Proportion of Statistical Significance Test (PSST; Stanley et al., 2021). PSST is the heterogeneous version of the Test of Excess Significance. It uses meta-analytic estimates of the mean effect size and between-study heterogeneity to compute the expected probability that each study will produce a statistically significant result. These expected probabilities are then averaged to obtain the expected proportion of significant findings. Publication bias is inferred when the observed proportion of significant findings is significantly higher than the expected proportion. Its limitation is that the expected discovery rate is estimated from the observed meta-analytic effect-size distribution, which may itself be distorted by publication bias or by misspecification of the heterogeneity model.
4. The Caliper Test (CT; Gerber & Malhotra, 2008a, 2008b). The caliper test compared the number of p-values just below and just above the significance threshold. The interval width was set to 20% of the criterion value, z = 1.96. This value was chosen because it was the widest interval evaluated in prior simulation work and showed only modestly inflated Type I error rates (Schneck, 2017). Narrower intervals provide a better local approximation around the threshold, but they also use fewer observations and therefore have lower power. Because the present study focuses on identifying competitive bias-detection methods under heterogeneous conditions, the wider caliper was used as the most favorable version of the test.

5. The step-function selection model developed by Vevea and Hedges (1995) and extended by Vevea and Woods (2005) was fitted with the R package weightr (Coburn & Vevea, 2019). I refer to the one-step specification as VW1 or 3PSM. The simulation also included Kranz and Kasy's implementation of a closely related model, but the results were virtually identical; therefore, only the results from weightr are reported.
6. Z-curve (Bartoš & Schimmack, 2022; Brunner & Schimmack, 2020). Z-curve estimates the expected discovery rate from the distribution of statistically significant z-values. For the bias test, bootstrap samples were drawn from the observed distribution, and both the observed discovery rate and expected discovery rate were recomputed in each bootstrap sample. The test statistic was D = ODR − EDR. Because publication bias predicts ODR > EDR, the one-sided bootstrap p-value was computed as the proportion of bootstrap samples in which $D \leq 0$. A low p-value therefore indicates that the observed discovery rate is reliably higher than the expected discovery rate.

## Results

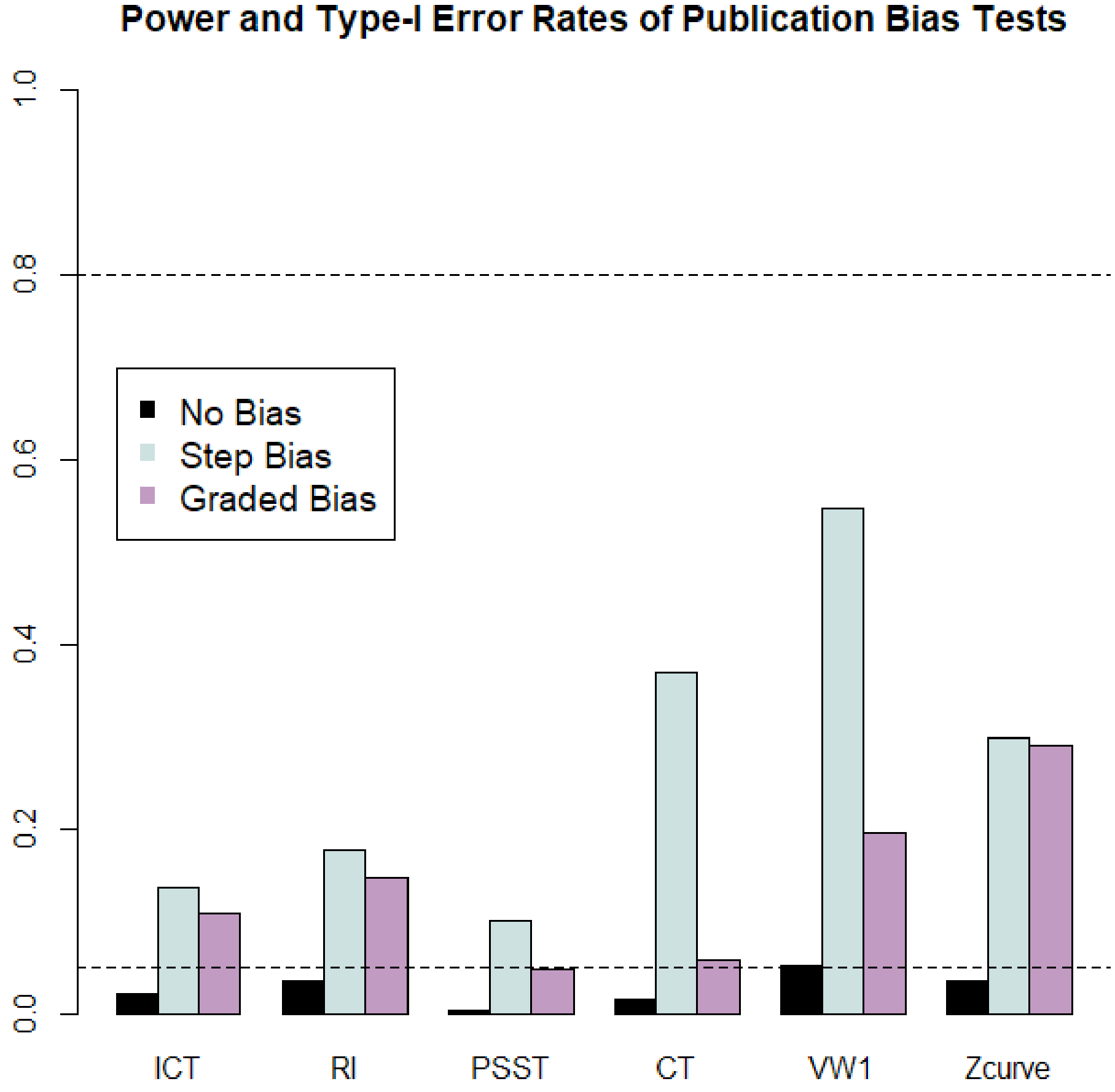

Figure 4

Note: ICT = Incredibility Test, RI = R-Index, PSST = Publication Selection for Significance Test, CT = Caliper Test, VW1 = Vevea-Woods 1 step selection model.

Figure 4 shows the overall Type 1 error rates and power estimates averaged over the 5,184 cells of the design. Under no publication bias, all methods produced low rejection rates, with most estimates close to or below the nominal .05 level. Thus, the simulations do not indicate severe overall inflation of Type 1 error. However, power to detect publication bias was generally modest. No method approached the conventional .80 benchmark when averaged across the full design.

The relative performance of the methods depended strongly on the form of selection. The one-step Vevea and Woods model had the highest power under step-function selection, as expected because the simulated selection mechanism matched the model assumption. The caliper test also performed better under step selection than under graded selection because step selection produces a clear drop in the number of nonsignificant results just below .05. In contrast, z-curve showed similar power under step and graded selection. This pattern is consistent with the fact that z-curve estimates bias from the discrepancy between the observed and expected discovery rates, rather than by modeling the shape of the nonsignificant distribution.

The main conclusion is that none of the tests demonstrated acceptable power to detect publication bias in simulations with at least 100 studies. A second important observation is that the two model-based approaches, the one-step Vevea and Woods model and z-curve, performed better overall than the other tests. The following analyses therefore focus on these two methods. Readers interested in the performance of the other tests under specific conditions can reproduce additional comparisons using the simulation code available on OSF.

**Sample Size and False Discovery Rates**

Table 1 shows the power of the two main bias tests as a function of the mixture of false-positive and true-positive results and the number of studies. The key result is again that power to detect publication bias is modest, even with k = 1,000 studies. The only condition in which power approached acceptable levels was the condition that matched the assumptions of the step-function model: a unimodal effect-size distribution, no false-positive mixture, and selection that followed a step function at the significance threshold. In this specific condition, the one-step Vevea and Woods model approached 80% power with k = 100 studies.

**Table 1**
*Type I Error Rates and Power for Bias Detection Tests by Bias Condition, Sample Size (k), and FDR*

| FDR | k | No Bias | | Step | | Graded | |
|---|---|---|---|---|---|---|---|
| | | VW1 | Z | VW1 | Z | VW1 | Z |
| 0.0 | 100 | 0.06 | 0.09 | 0.77 | 0.43 | 0.37 | 0.46 |
| 0.0 | 300 | 0.06 | 0.10 | 0.91 | 0.59 | 0.53 | 0.62 |
| 0.0 | 1000 | 0.03 | 0.08 | 0.95 | 0.83 | 0.67 | 0.74 |
| 0.2 | 100 | 0.06 | 0.02 | 0.38 | 0.11 | 0.12 | 0.11 |
| 0.2 | 300 | 0.06 | 0.01 | 0.56 | 0.19 | 0.10 | 0.20 |
| 0.2 | 1000 | 0.01 | 0.03 | 0.58 | 0.41 | 0.21 | 0.33 |
| 0.4 | 100 | 0.06 | 0.04 | 0.36 | 0.11 | 0.07 | 0.05 |
| 0.4 | 300 | 0.05 | 0.01 | 0.38 | 0.15 | 0.05 | 0.15 |
| 0.4 | 1000 | 0.04 | 0.01 | 0.43 | 0.31 | 0.10 | 0.27 |
| 0.6 | 100 | 0.04 | 0.00 | 0.37 | 0.07 | 0.05 | 0.06 |
| 0.6 | 300 | 0.07 | 0.00 | 0.49 | 0.16 | 0.03 | 0.13 |
| 0.6 | 1000 | 0.08 | 0.02 | 0.52 | 0.34 | 0.07 | 0.33 |

Note. VW1 = Vevea-Woods 1 step (.025); Z = z-curve test. FDR = false discovery rate. No Bias, Step, and Graded refer to the selection model under which tests were evaluated.

This result is noteworthy because Hedges and Vevea reported much lower power for selection models. One possible reason is that their model included multiple selection steps, which may reduce power when the true selection process contains only a single step. A one-step model works well when selection bias is also generated by a single step. However, the results for graded selection show that power decreases considerably when this assumption is violated.

The second important finding is that z-curve's relative advantage under graded selection does not translate into high power to detect publication bias. Even with k = 1,000 studies, power remains modest. Thus, the main problem is not that one method uniformly dominates another. Rather, bias detection in meta-analysis is highly conditional on model assumptions and often has high Type II error rates. In many realistic conditions, bias tests fail to detect publication bias even when 70% of nonsignificant results are missing.

### Visible Moderators

A limitation of moderator analyses based on simulated factors is that many of the moderators are not directly observable in applied data. In contrast, visual inspection of z-curve plots can provide useful information about the likely power of bias tests, even before formal estimates are considered (Bartoš & Schimmack, 2026). Z-curve relies only on significant results, whereas the main simulation design varied the total number of studies because other methods use both significant and nonsignificant results. As a result, z-curve sometimes had relatively few significant results from which to estimate the EDR. It is therefore useful to examine z-curve's performance as a function of the absolute number of significant results.

The second visible moderator is the shape of the significant z-value distribution. The most relevant feature is the slope of the histogram between $z = 2$ and $z = 3$, which can be estimated by regressing bin densities on z-values in this range. A negative slope indicates an excess of just-significant results relative to stronger significant results. This pattern implies low average power and predicts many nonsignificant results under the fitted model. Consequently, a negative slope should make publication bias easier to detect than a flat or positive slope, provided that the number of significant results is sufficiently large. This condition is also substantively important because selection for significance produces the greatest inflation of effect-size estimates when studies have low power.

Figure 5 shows power as a function of these two visible moderators: the number of significant results and the slope of the significant z-value distribution. The main result is again that power remains modest in most conditions. However, when the number of significant results exceeded 300 and the significant z-value distribution had a negative slope, z-curve approached the conventional criterion of acceptable power. In this condition, z-curve outperformed the step-function model, especially when selection bias was not uniform across the nonsignificant range.

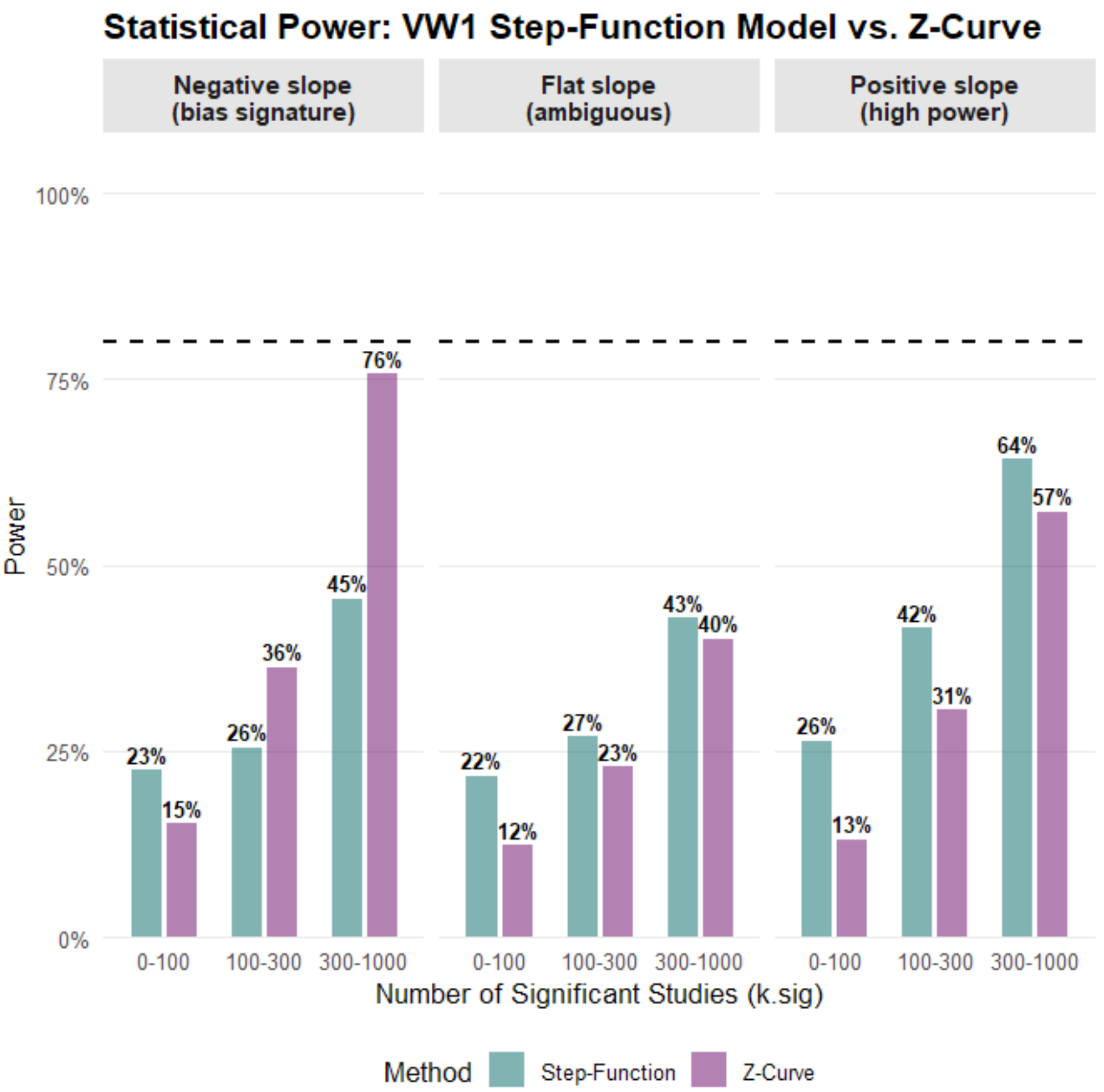

These results provide a practical diagnostic for applied use. A nonsignificant z-curve bias test is difficult to interpret when the number of significant results is small or when the significant z-value distribution is flat. In contrast, when many significant results are available and the z-curve plot shows a clear excess of just-significant results, the bias test has substantially more diagnostic value. Thus, the z-curve plot provides visible information about when publication-bias detection is likely to be informative and when estimation with wide confidence intervals is the more appropriate interpretation. Z-curve plots are also useful for selection models. Even though the input are effect size estimates and standard errors, the selection bias is modeled as a function of p-values or corresponding z-values. Z-curve plots make it possible to visually inspect the relevant information directly, while funnel plots provide the same information only indirectly.

**Estimation of Publication Bias**

The previous results showed that bias tests often have low power in realistic settings when the distribution of effect sizes is unknown, even when the number of studies is in the hundreds. Although this finding is not new (Hedges & Vevea, 1996; Renkewitz & Keiner, 2019), it has been neglected in applied meta-analysis, where nonsignificant bias tests are often reported as if they provided evidence against publication bias. This interpretation is problematic because bias tests frequently have high Type II error rates.

The problem has also been obscured by the distinction between bias detection and bias correction. A model that fails to detect bias may also fail to correct it, and a model that detects model misfit may still produce a misleading correction (Figure 2). Thus, the central question is not simply whether a bias test rejects the null hypothesis. It is therefore necessary to take a new approach to the problem of publication bias in meta-analyses.

A solution to this problem is to focus on effect size estimation with confidence intervals that quantify uncertainty in the amount of bias. The key difference is that nonsignificant results provide information for worst-case scenarios. Even if a confidence interval includes zero and is consistent with the absence of bias, the upper bound may be high and alert readers to the fact that bias could also be large. Thus, even nonsignificant results are informative. When confidence intervals are wide, they reveal that publication bias can be large. The approach also makes it possible to distinguish between meta-analysis with informative and uninformative data. Informative meta-analyses have small confidence intervals that rule out substantial bias.

Step-function models and z-curve measure the amount of bias differently. Step-function models estimate a weight parameter (w) for nonsignificant results relative to a weight of 1 for significant results. Bias is therefore $1 - w$. Z-curve estimates bias as the discrepancy between the observed discovery rate and the expected discovery rate (ODR − EDR). To facilitate direct comparison, z-curve EDR estimates were converted to w using the following formula, so that both models are evaluated on the same estimand — the publication probability for nonsignificant results:

$$w = EDR(1 - ODR) / (ODR(1 - EDR))$$

In all simulations, the target selection mechanism retained 30% and dropped 70% of nonsignificant results. However, other constraints made it impossible to match this value exactly in every simulation. The median bias was 64% for step selection and 63% for graded selection. For the model comparison, only runs with true bias between 40% and 80% were retained. This excluded 2% of simulations from the analysis.

Figure 6 shows the results. Z-curve confidence intervals included the true value 97% of the time, slightly above the nominal 95% level. In contrast, the upper bound of the confidence interval for the step-function model was often below the true value of bias, and only about half of the confidence intervals included the true value. Thus, z-curve performed well in absolute terms and substantially better than the step-function model in estimating the amount of selection.

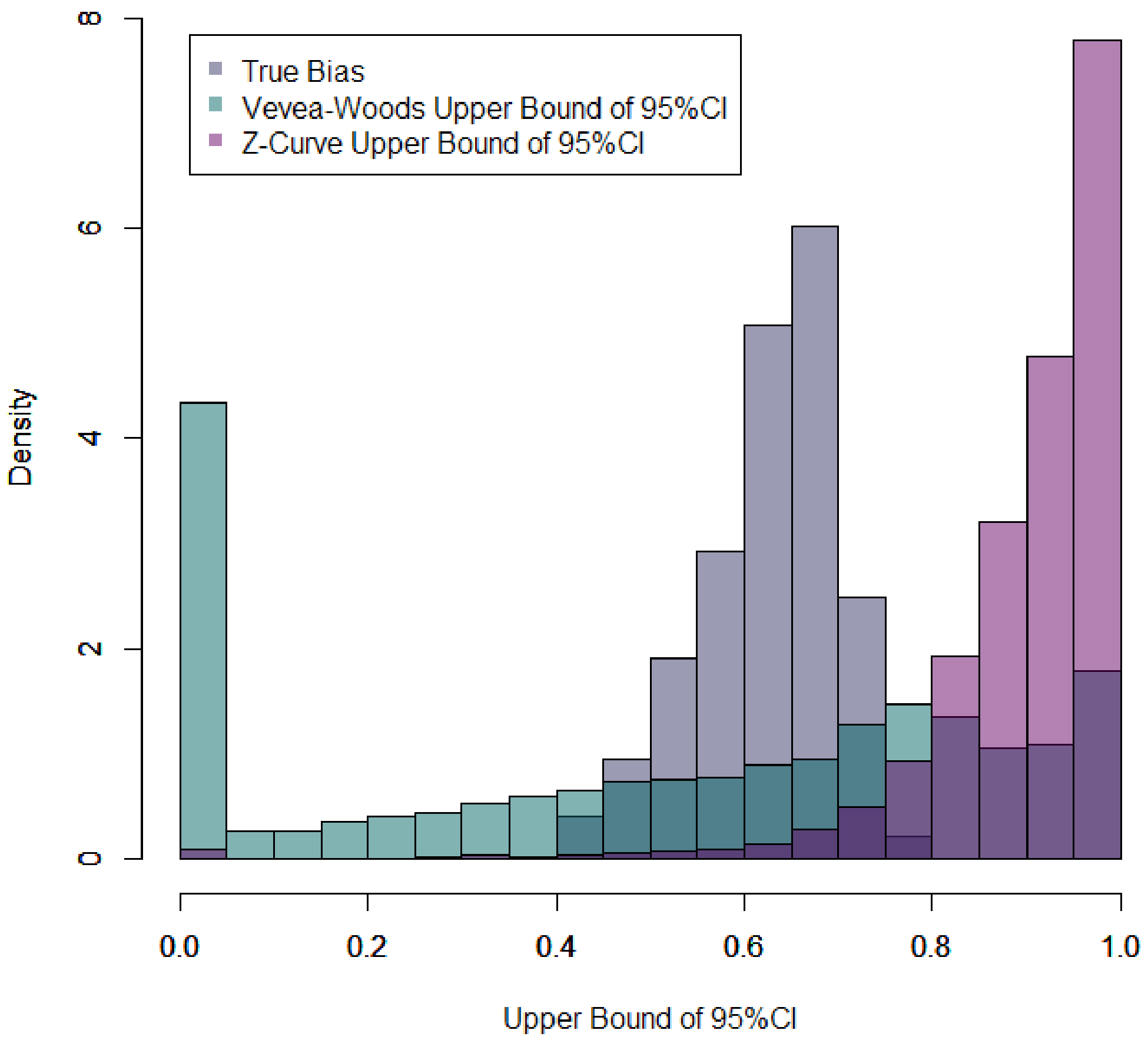

Figure 6

Upper Bound of the Confidence Interval for the Percentage of Missing Nonsignificant Results

The relatively poor performance of the step-function model compared with Hedges and Vevea's results can be explained by greater heterogeneity in the present simulations. In simulations with small heterogeneity (tau ≤ .2), no false-positive results (FDR = 0), and uniform selection bias, coverage of the 95% confidence interval was 94%. This finding supports Hedges and Vevea's conclusion that the model works well for meta-analyses of direct replication studies with small heterogeneity. At the same time, the full set of results confirms their conclusion that performance decreases when heterogeneity is larger and model assumptions are violated. Thus, the present results extend rather than contradict Hedges and Vevea's (1996) robustness test. In difficult situations with unknown selection bias, an unknown distribution of population effect sizes, and substantial heterogeneity, z-curve is preferable because it requires fewer assumptions.

**Illustration 1: Meta-Analysis of Social Priming (Dai et al., 2023)**

Social priming without awareness was an influential paradigm in experimental social psychology in the 1990s and 2000s, until the replication crisis cast doubt on the robustness of these findings (Kahneman, 2012, 2017; Pashler et al., 2012; Schimmack et al., 2017; Shanks et al., 2013; Sotola, 2022). Meta-analyses show that older articles reported moderate effect size estimates, but newer preregistered studies show no evidence of an effect (Dai et al., 2023). This makes social priming a prominent candidate to evaluate publication bias tests with real data.

Dai et al. (2023) compiled a large dataset with k = 862 statistical results from 351 independent studies. They analyzed these data with the step-function selection model of Vevea and Woods (2005) and reported a corrected estimate of the average effect size of d = .27. However, they did not report a confidence interval around this point estimate or the estimated heterogeneity parameter for population effect sizes. Moreover, their selection-model analysis did not take the nested structure of the data into account because the R package assumes independent effect sizes.

I reanalyzed their data, excluding six effect sizes larger than d = 2.5, with a clustered-bootstrap sampling approach to obtain a confidence interval around the weightr estimates. With a focus on positive effects, it is preferable to specify a step at p = 0.50 (one-tailed). This way the model does not require untestable assumptions about biases in reporting of reversed priming effects. This specification also limits the width of the step for nonsignificant results in the expected direction and makes the model more robust.

The estimated population mean was similar to the one reported by Dai et al. (2023), mu = .22, 95%CI [.19 , .24]. The estimated heterogeneity was tau = .38, 95%CI [.33, .44]. The weight parameter for nonsignificant positive results was .62, 95%CI [.51, .73]. Thus, the upper limit for publication bias is 1-.51 = 49%. However, the large heterogeneity and the simulation results raise concerns that the confidence interval around the selection-weight estimate may be too narrow.

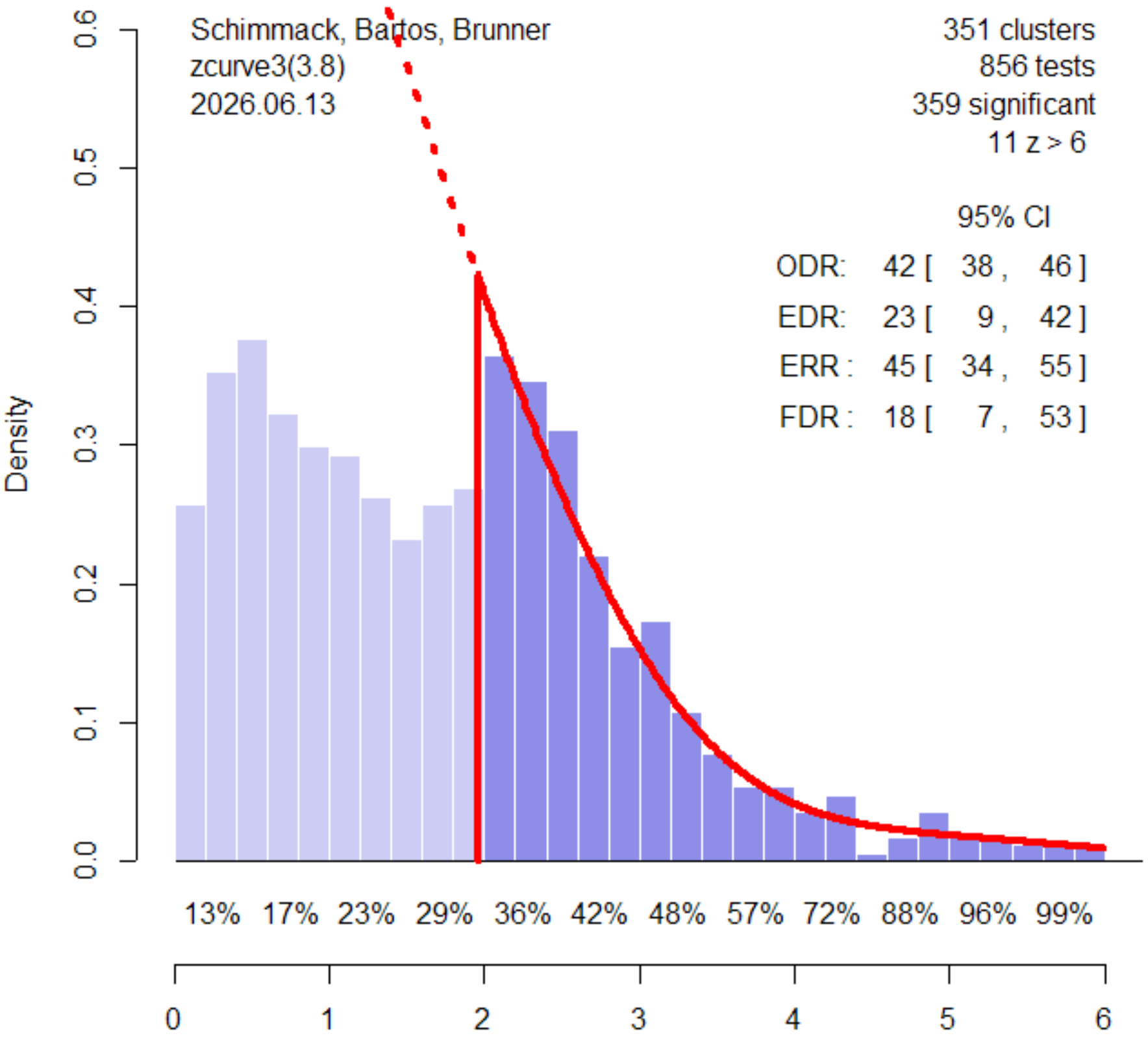

Figure 7

Z-Curve Plot of Social Priming Meta-Analysis by Dai et al. (2023)

Figure 7 shows the z-curve plot. The EDR estimate is 23%, 95% CI [9%, 42%]. This translates into an estimated selection weight of 41%, 95% CI [14%, 100%]. Thus, the upper confidence limit for publication bias, defined as 1 minus the selection weight, is 1 − .14 = .86. This is much higher than the estimate obtained from the step-function selection model.

The z-curve estimate of selection bias can also be used to obtain confidence intervals for the average effect size and heterogeneity (μ & τ). For each z-curve EDR estimate and confidence limit, I converted EDR to the equivalent selection weight and fitted a fixed-weight

Vevea-Woods model in weightr; z-curve therefore supplied only the selection-weight scenarios, whereas $\mu$ and $\tau$ were estimated by the standard selection model. The 95% CI for the average effect size ranges from d = .07 to d = .37. The 95% CI for $\tau$ ranges from .30 to .38. Using the worst-case scenario compatible with the data, the prediction interval ranges from .07 − 2(.38) to .07 + 2(.38), or from d = −.69 to d = .83. In other words, without further information, the population effect size for a new priming study could range from a moderate negative effect to a large positive effect. Thus, the data provide very little information about the effect size of any specific priming manipulation, even though the point estimate for the average effect size is significantly different from zero.

Z-curve also provides information that conventional effect-size meta-analyses do not provide. High heterogeneity makes it possible that some significant results are false positives. Z-curve uses the EDR estimate to quantify the maximum false discovery rate using Sorić's (1989) formula. The point estimate of the EDR is compatible with up to 18% false positives. This risk increases to 53% when the lower bound of the EDR confidence interval is used.

Optimists may treat this finding as a glass-half-full result because even the lower-bound estimate implies that roughly half of the significant results are not false positives. However, this conclusion is of limited practical value because it remains unclear which studies are true positives and which are false positives. Moreover, true positives can still have very small effect sizes. This helps explain why it has been difficult to find original social-priming results that replicate, even with larger samples (Dai et al., 2023).

In conclusion, Dai et al.'s conclusions about social priming based on the selection-model results do not hold upon further inspection. Although the point estimate suggests a small average effect, this estimate does not adequately summarize the literature. Uncertainty about the amount of publication bias makes the average-effect estimate unreliable, and there is substantial heterogeneity in population effect sizes. The uncorrected effect-size estimate highlighted in Dai et al.'s abstract, d = 0.37, is therefore misleading as a summary of the social-priming literature. It overstates robustness in three ways. First, the estimate is inflated by publication bias. Second, by featuring the point estimate without its confidence interval, it conceals the fact that substantially smaller average effects remain compatible with the data. Third, substantial heterogeneity implies that even a positive average effect provides little information about the direction or magnitude of the effect for any specific priming manipulation.

**Illustration 2: Meta-Meta-Analysis of Applied Psychology (Siegel et al. 2022)**

Siegel et al. (2022) re-examined 74 meta-analyses published in the Journal of Applied Psychology with at least 10 effect sizes (observed range 11 to 325; median = 37). Running all six bias tests on these data produced very low detection rates. The step-function model had the highest rate with 11%. In other words, 89% of the time the results were inconclusive. Either the meta-analysis was unbiased, or the tests failed to detect bias.

Focusing on estimation of bias with confidence intervals is more informative. The step-function model's average weight for nonsignificant results was 1.43, 95%CI [1.09, 1.77], median = 0.94, indicating no substantial publication bias on average. Thus, the low percentage of nonsignificant tests is mostly explained by the absence of substantial selection bias. This is confirmed by the z-curve results. Z-curve shows a small average bias, ODR – EDR = .07, 95%CI [.00, .14].

These results underscore the value of estimation over nil-hypothesis testing. Nonsignificant results of bias tests are inconclusive and cannot assure readers that publication bias is absent. In contrast, evidence that publication bias is small with tight confidence intervals suggests that the main results are credible and not substantially inflated.

I also examined two specific meta-analyses that were flagged as biased by 4 of the 6 bias tests. Figure 8 shows the z-curve plots for these two meta-analyses.

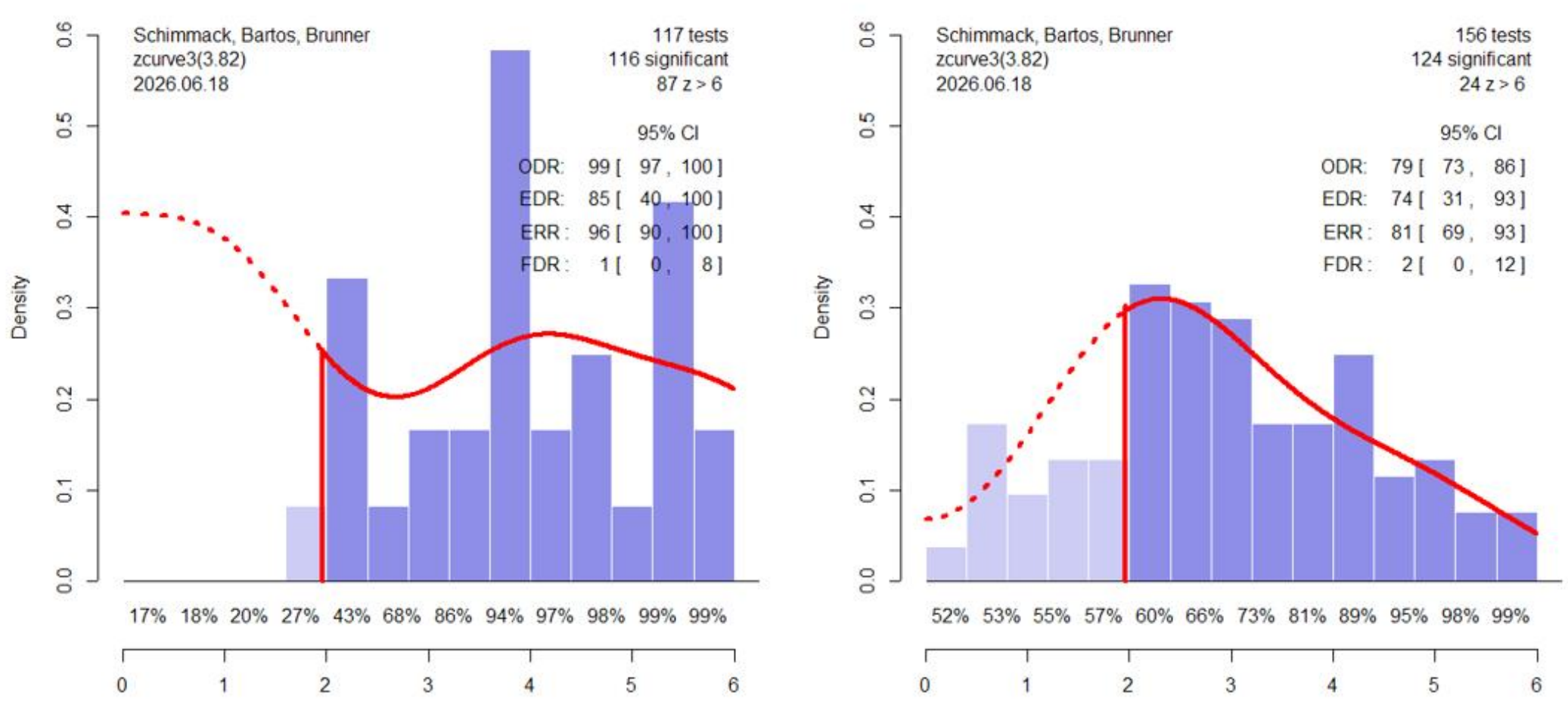

Figure 8. Z-curve plots of two meta-analyses from Siegel et al. (2022).

Bias was detected because both meta-analyses were relatively large (k > 100) and most studies had high power, making it possible to detect the absence of a few nonsignificant results. However, the amount of bias is small and had no practical implications for the estimation of the average effect size. These results further underscore the problem of publication bias detection methods. They often fail to detect bias and when they detect it, it may be inconsequential. The solution to this problem is estimation.

**Discussion**

Publication bias tests have a long history and are routinely used in meta-analyses to test the robustness of effect size estimates to violations of the assumption that the available data are a representative set of studies that were conducted. In recent years, statisticians have also developed new methods to correct effect size estimates for inflation due to selection for significance. These methods are increasingly used to assure readers that meta-analytic results are credible (Dai et al., 2023). One of the few evaluation studies showed that bias tests have low power when heterogeneity is high and the set of studies is small (Renkewitz & Keiner, 2019). The present simulation studies showed that the problem persists even in large meta-analysis with hundreds of effect sizes, unless the data mirror the assumptions of bias tests. The main problem is structural. Detecting publication bias is difficult because bias introduces systematic error that is not minimized by increasing sample sizes (Hedges & Vevea, 1996).

**Moving From Detection to Estimation**

While detection of publication bias is often out of reach when the number of studies is not large and some nonsignificant results are reported, it is still necessary to evaluate the influence of publication bias on meta-analytic estimates of effect sizes. The solution is to quantify the amount of bias with confidence intervals. In small sets of studies, these confidence intervals are often wide. Wide confidence intervals are often seen as a problem and used to argue that a test is useless (McShane et al., 2020). This is true for original research where it is possible to reduce the width of confidence intervals by increasing sample sizes. However, the number of studies in a meta-analysis at any moment in time is fixed. In this case, wide confidence intervals are still informative because they specify the range of values that are consistent with the data (Soto & Schimmack, 2026). Thus, the upper bound of the confidence interval for bias can be used to examine the implications of this amount of bias for the conclusions. If bias is large and the main claim is not robust to this amount of bias, the data provide no evidence for this claim. Moreover, uncertainty about the amount of bias increases uncertainty about the true population effect size. The wider confidence intervals around the effect size estimate now properly reflect the uncertainty in these estimates that is created by uncertainty about the amount of bias. The onus is then on researchers to conduct new studies to provide stronger evidence for their claims. This new way of thinking about statistical results is slowly being adapted for original research. It is time to do the same for bias tests and meta-analytic effect size estimation.

**Estimation with Calibrated Confidence Intervals**

Confidence intervals are increasingly used to report results, but most of the discussion continues to focus on the noisy point estimate. When the confidence interval does not include zero, discussion often shifts to the point estimate. This is not justified because other values within the confidence interval are consistent with the data at the same error level. To avoid this fallacy, it is important to focus on the boundaries of the confidence interval. If the point estimate of bias is .1 (a difference of 10 percentage points between the observed and expected discovery rate), but the upper bound of the confidence interval is .5, it is not justified to claim that bias is small. Instead, large bias remains consistent with the data. Thus, discussions of bias need to focus on the upper limit. This is akin to sensitivity analyses for publication bias, but there is an important difference. Sensitivity analyses rely on assumptions and can be dismissed as making implausible assumptions (Dai et al., 2023). In contrast, confidence intervals are a range of values based on the same assumptions and it is not possible to argue that some values outside of the interval are implausible. The fact that these values are included in the confidence interval implies that they are plausible values, even if the point estimate is small.

**Coverage of Confidence Intervals**

Confidence intervals express uncertainty about the results of a study in terms of long-run properties of a statistical procedure and error rates. A 95% confidence interval promises to contain the true value 95% of the time, but 5% of the time the true value may be outside the interval. This information is only useful if the true error rate matches the nominal one. Simulation studies are needed to test the behavior of bias tests, just like bathroom scales undergo quality control to ensure that the stated margin of error holds across many uses of the same product. However, many bias tests lack robustness tests under a broad range of conditions or are used even if their power has been shown to be low (Renkewitz & Keiner, 2019). Here I showed that the step-function model works only under restrictive assumptions that match the model's assumption about the shape of bias and distribution of effect sizes, but not under robust realistic conditions. I also showed that z-curve has much better coverage and meets nominal coverage claims because z-curve has been validated with extensive simulation studies (Bartoš & Schimmack, 2022).

The better coverage of z-curve also implies that z-curve provides better information about uncertainty in effect size estimates. Step-function models' confidence intervals around the average effect size estimate are too narrow because they underestimate the upper limit of bias consistent with the data. By converting z-curve estimates of the EDR into estimates of the weights for step-function models, z-curve can be used to obtain confidence intervals for effect size estimates with better coverage.

### Estimation Versus Sensitivity Analyses

The focus on estimation with confidence intervals is closely related to sensitivity analysis (Vevea & Woods, 2005). Both acknowledge uncertainty about the true amount of publication bias and ask whether substantive conclusions are robust to more bias than the point estimate suggests.

Vevea and Woods (2005) proposed sensitivity analysis for the common situation in which a meta-analysis contains too little information to estimate the selection weights. In this case, researchers fix the weights at assumed values and estimate the average effect size and heterogeneity conditional on those weights. The limitation is that the assumed weights are chosen by the analyst. Conclusions therefore depend on the analyst's specification, and an unfavorable result can be dismissed as resting on unrealistically severe assumptions (Dai et al., 2023).

The present approach uses the lower bound of z-curve's EDR confidence interval as the most severe scenario for publication bias. This worst case is still hypothetical, but it is based on actual data and for this reason cannot be dismissed as speculative and unrealistic. The width of the interval continues to matter, but in the direction Vevea and Woods anticipated: small meta-analyses contain little information about selection and therefore cannot rule out large amounts of bias. This undermines the point estimate, but not the upper bound, which correctly reports that substantial bias remains compatible with the data and that uncertainty about bias adds to uncertainty about the population effect size. Using this empirically fixed weight thus serves the purpose of a sensitivity analysis without introducing a subjective specification.

In this approach, z-curve provides the selection weights and the step-function model provides estimates of the mean and heterogeneity of population effect sizes. Replacing the estimated weights from the step-function model with those based on z-curve addresses the problem that the step-function model often underestimates uncertainty (Figure 6). The key limitation is that estimates of the mean and heterogeneity of the population effect sizes still assume a normal distribution, but the estimates no longer make assumptions about the shape of selection bias.

**Limitations**

All models work well when their assumptions are met and may fail when they are violated; z-curve is no exception. The main assumption underlying z-curve estimates of the EDR is that selection bias is primarily a function of statistical significance, or equivalently, of a study's probability of producing a significant result. The selection assumption can be violated when researchers use questionable research practices to increase the probability of obtaining significance (John et al., 2012). Many of these practices produce just-significant p-values, leading to a bunching of z-values close to the significance criterion and a lower EDR estimate. In this case, z-curve may overestimate the severity of selection bias. This direction of error is conservative for the present purpose, however, because a lower EDR raises the upper bound on the amount of bias that remains compatible with the data. Z-curve does not identify the mechanism behind the excess of just-significant results; questionable research practices and a file drawer of nonsignificant results both lower the EDR, and for the purpose of bounding the credibility of a literature this is the desired behavior.

A more serious concern, because it operates in the opposite direction, is selection that leads z-curve to overestimate the EDR and thereby underestimate bias. This can occur when significant results with larger effect sizes are reported in preference to those with smaller effect sizes. One simulation study suggested that this bias may inflate EDR estimates up to 10 percentage points (Pek et al., 2026; Schimmack & Soto, 2026). This should be considered an upper limit because effect size selection is unlikely to occur without questionable research practices that have the opposite effect. Further work on selection among significant results is needed, but it bears equally on step-function models. Although additional steps within the significant range can be specified, applications and previous simulation studies have assumed that significant results are observed without further selection (Carter et al., 2019; Hedges & Vevea, 1996).

**Conclusion**

The main contribution of this article is to shift the goal of meta-analysis from detecting and correcting publication bias to estimating how much bias remains compatible with the data, especially in literatures with high heterogeneity. Previous work has tried to detect bias, but these methods have low statistical power even with large sets of studies.

The more defensible approach is to estimate bias with confidence intervals. Step-function selection models are poorly suited to this purpose when their assumptions are not met because their intervals are too narrow and understate the amount of bias. Z-curve's EDR confidence intervals have better coverage under these conditions and can be used to quantify bias.

Step-function models can also be used for sensitivity analyses, but it is unclear how much bias should be assumed for a given dataset. Using the confidence interval of z-curve’s EDR estimates solves this problem and can be used to obtain adjusted estimates of the average and heterogeneity of effect sizes in a study. In short, z-curve is a valuable tool for meta-analyses of conceptual replication studies with large heterogeneity and publication bias.